\documentclass[prl,superscriptaddress,twocolumn,letterpaper,amssymb,amsmath,floatfix]{revtex4-1}
\pdfoutput=1
\usepackage{gensymb}
\usepackage{amsmath}
\usepackage{color}
\usepackage{graphicx}
\usepackage{mathrsfs}
\usepackage{youngtab}
\usepackage{epstopdf}

\begin{document}

\title{Modular matrices from universal wave function overlaps \\
 in Gutzwiller-projected parton wave functions}

\author{Jia-Wei Mei}
\affiliation{Perimeter Institute for Theoretical Physics, Waterloo, Ontario, N2L 2Y5 Canada}
\author{Xiao-Gang Wen}
\affiliation{Department of Physics, Massachusetts Institute of Technology, Cambridge, Massachusetts 02139, USA}
\affiliation{Perimeter Institute for Theoretical Physics, Waterloo, Ontario, N2L 2Y5 Canada}
\date{\today}
\begin{abstract}
We implement the universal wave function overlap (UWFO) method to extract
modular $S$ and $T$ matrices for topological orders in Gutzwiller-projected
parton wave functions (GPWFs). The modular $S$ and $T$ matrices generate a
projective representation of $SL(2,\mathbb{Z})$ on the degenerate-ground-state
Hilbert space on a torus and may fully characterize the 2+1D topological
orders, i.e. the quasi-particle statistics and chiral central charge (up to
$E_8$ bosonic quantum Hall states).  We used the variational Monte Carlo method
to computed the  $S$ and $T$ matrices of the chiral spin liquid (CSL)
constructed by the GPWF on the square lattice, and confirm that
the CSL carries the same topological order as the  $\nu=\frac{1}{2}$ bosonic
Laughlin state. We find that the non-universal exponents in UWFO can be small
and direct numerical computation is able to be applied on relatively large
systems. We also discuss the UWFO method for GPWFs on other Bravais lattices in
two and three dimensions by using the Monte Carlo method. UWFO may be a
powerful method to calculate the topological order in GPWFs.

\end{abstract}
\maketitle

Topological order\cite{Wen1990a,Wen1989,Wen1990} connotes the pattern of long-range entanglement in gapped
many-body wave functions\cite{Levin2006,Kitaev2006,Chen2010}. It describes
gapped quantum phases of matter that lie beyond the Landau symmetry breaking
paradigm\cite{Wen2004}. Local unitary transformations on many-body wave
functions can remove local entanglement, however, preserve the long-range
topological entanglement. Therefore, a topological ordered state is not
smoothly connected to a trivial (direct product) state by local unitary
transformations\cite{Chen2010}. Physically, topological order is described
through topological quantum numbers, such as non-trivial ground state
structures  and fractional
excitations.\cite{Laughlin1983,Wilczek1984,Arovas1984,Wen1989,Wen1990,Wen1990a}
These topological properties are fully characterized by the quasi-particle
(anyon in the bulk) statistics\cite{Laughlin1983,Wilczek1984,Arovas1984} and
the chiral central charge which encodes information about chiral gapless edge
states\cite{Wen1992,Wen1995}.

Both the fusion rule and the topological spin of quasi-particles as well as the
chiral central charge are characterized in the non-Abelian geometric phases
encoded in the degenerate ground
states\cite{Keski-Vakkuri1993,Wen1990a,Wen1989,Wen1990,Wen2012,Liu2013,Moradi2014,He2014},
and vise versa.  The non-Abelian geometric phases form a representation of
$SL(2,\mathbb{Z})$, that is generated by 90\degree rotation and Dehn twist on a
torus, which are called modular $S$ and $T$ matrices,
respectively\cite{Keski-Vakkuri1993,Wen2012}. The element of the modular $S$
matrix determines the mutual statistics of quasi-particles while the element of
the $T$ matrix determines the topological spin $\theta_a \in U(1)$ and the
chiral central charge\cite{Keski-Vakkuri1993,Wen1990a,Wen2012}.  

Given the fusion coefficients $N_c^{ab}$ and the topological spin $\theta_a$,
we can write down the modular $S$ and $T$ matrices as the following
expressions, $S_{ab} = \frac{1}{\mathcal{D}} \sum_cN_c^{a\bar{b}}
\frac{\theta_c}{\theta_a\theta_b}d_c$ and $T_{ab}=
e^{-i\frac{2\pi c}{24}}
\theta_a\delta_{a,b}$.\cite{Wang2010}  Here $d_a$ (called the quantum dimension
of quasiparticle $a$) is the largest eigenvalue of matrix $N_a$ which is
defined as $(N_a)_{bc}=N_c^{ab}$  and $\mathcal{D}$ is the total quantum
dimension, $\mathcal{D}^2=\sum_ad_a^2$.  We see that
$S_{a1}=\frac{d_a}{\mathcal{D}}$.

From Verlinde formula\cite{Verlinde1988}, we can reconstruct the fusion
coefficients, $N_{ab}^c=\sum_x\frac{S_{ax}S_{bx}S_{cx}^*}{S_{1x}}$. Therefore,
$S$ and $T$ provide a complete desciption and can be taken as the order
parameter of topological orders\cite{Wen2012,Liu2013,Moradi2014,He2014}.The
modular $S$ and $T$ matrices satisfy the relations,  $(ST)^3=
C$ and $S^2=C$, where $C$ is a so-called charge conjugation matrix that
satisfies $C^2=1$. The central charge $c$ determines the thermal current of the
edge state, $I_E =\frac{c}{6} T^2$, at temperature $T$\cite{Affleck1986} and is
fixed up to $E_8$ bosonic quantum Hall states.

To fully  characterize topological order, various numerical methods are
proposed to access the modular $S$ and $T$
matrices\cite{Zhang2012,Cincio2013,Tu2013,Zaletel2013,Zhu2013,Zhang2014}.
Recently, one of us proposed the universal wave function overlap (UWFO) method
to calculate modular matrices\cite{Hung2014,Moradi2014}. For a given set
$\{|\psi_a\rangle\}_{a=1}^N$ of degenerate ground-state wave functions, it
provides us a practical method to extract the modular $S$ and $T$ matrices
\begin{eqnarray}
  \label{eq:uwfo}
  \tilde{S}_{ab}=\langle\psi_a|\hat{S}|\psi_b\rangle&=&e^{-\alpha_S L^2 + o(1/L^2)}S_{ab},\nonumber\\
  \tilde{T}_{ab}=\langle\psi_a|\hat{T}|\psi_b\rangle&=&e^{-\alpha_T L^2 + o(1/L^2)}T_{ab},
\end{eqnarray}
where $\hat{S}$ and $\hat{T}$ are the operators that generate the
90\degree~rotation and Dehn twist, respectively, on a torus with the $L^2$
lattice size. The exponentially small prefactor makes it difficult to
numerically calculate the  UWFO in (\ref{eq:uwfo}). To avoid the exponential
smallness, a gauge-symmetry preserved tensor renormalization method has been
developed for the tensor-network wave functions\cite{Moradi2014,He2014}, where
the system size is effectively reduced as zero after the tensor
renormalization.

Actually, in this letter, we will show that the non-universal  exponent
$\alpha_{T,S}$ can be small such that the UWFO can be directly numerically
calculated on relatively large systems. We will take a chiral spin liquid (CSL)
wave function on the square lattice\cite{Wen1989a} as an explicit example to
extract the modular $S$ and $T$ matrices from the UWFO. We construct the set of
the ground states for a CSL by using Gutzwiller-projected parton wave functions
(GPWF).\cite{Kalmeyer1987,Wen1991,Wen1989a,Wen1999,Zhang2012} We use the
variational Monte Carlo to calculate the UWFO for the CSL wave functions. The
hopping parameters are set as $|t_1/t_0|=0.5$ for the CSL on the $\pi$-flux square lattice,
 where $t_0$ and $t_1$ for nearest neighbor and next nearest
neighbor links, respectively. Since $C_4$ symmetry, the overlap $\tilde{S}$ in
Eq. (\ref{eq:uwfo})  has a vanishing exponent $\alpha_S = 0$. $\tilde{T}$ in
Eq. (\ref{eq:uwfo}) has the relatively small non-universal complex exponent
$\alpha_T=0.04208+0.07654i$ and the direct numeric computation is carried out
on relatively large systems up to $12\times12$ lattice size in this letter. The
CSL is the lattice analogy of $\nu=\frac{1}{2}$ bosonic Laughlin
state\cite{Kalmeyer1987,Wen1989a}. Our numerical results confirm the analogy by
directly extracting the modular $S$ and $T$ matrices from the UWFO.

In the parton construction,  the $S=\frac{1}{2}$ spin operator is  written in
terms of fermionic parton operators, $ S^a(z_i)=\frac{1}{2}f_{\sigma}^\dag(z_i)
\sigma_{\sigma\sigma'}^a f_{\sigma'}(z_i)$. Here $\sigma^a$ $(a=x,y,z)$ is the
Pauli matrices and $f_{i\sigma}$ ($\sigma=\uparrow/\downarrow$) is the
fermionic parton operator.  We take the complex variables for the $i$-site
coordinate, $z_i=x_i+iy_i$, on a lattice. We have to impose the
one-particle-per-site constraint for the partons, $f_{\uparrow}^\dag(z_i)
f_{\uparrow}(z_i)+f_{\downarrow}^\dag(z_i) f_{\downarrow}(z_i)=1$, such that
the fermionic partons have the same Hilbert space on $i$-site as the spin
operators $S^a(z_i)$. The GPWF for the spin system can be read as
\begin{eqnarray}
\label{eq:gpwf} |\Psi\rangle=\sum_{\{z_i\}}\mathcal{P}_G\Psi(\{z_i^\uparrow,z_k^\downarrow\})|\{z_i\}\rangle,
\end{eqnarray}
where $|\{z_i\}\rangle$ the spin configuration and  $\mathcal{P}_G$ is the Gutzwiller projection operator to impose the one-particle-per-site constraint for the fermionic partons.

\begin{figure}[t]
  \centering
  \includegraphics[width=0.8\columnwidth]{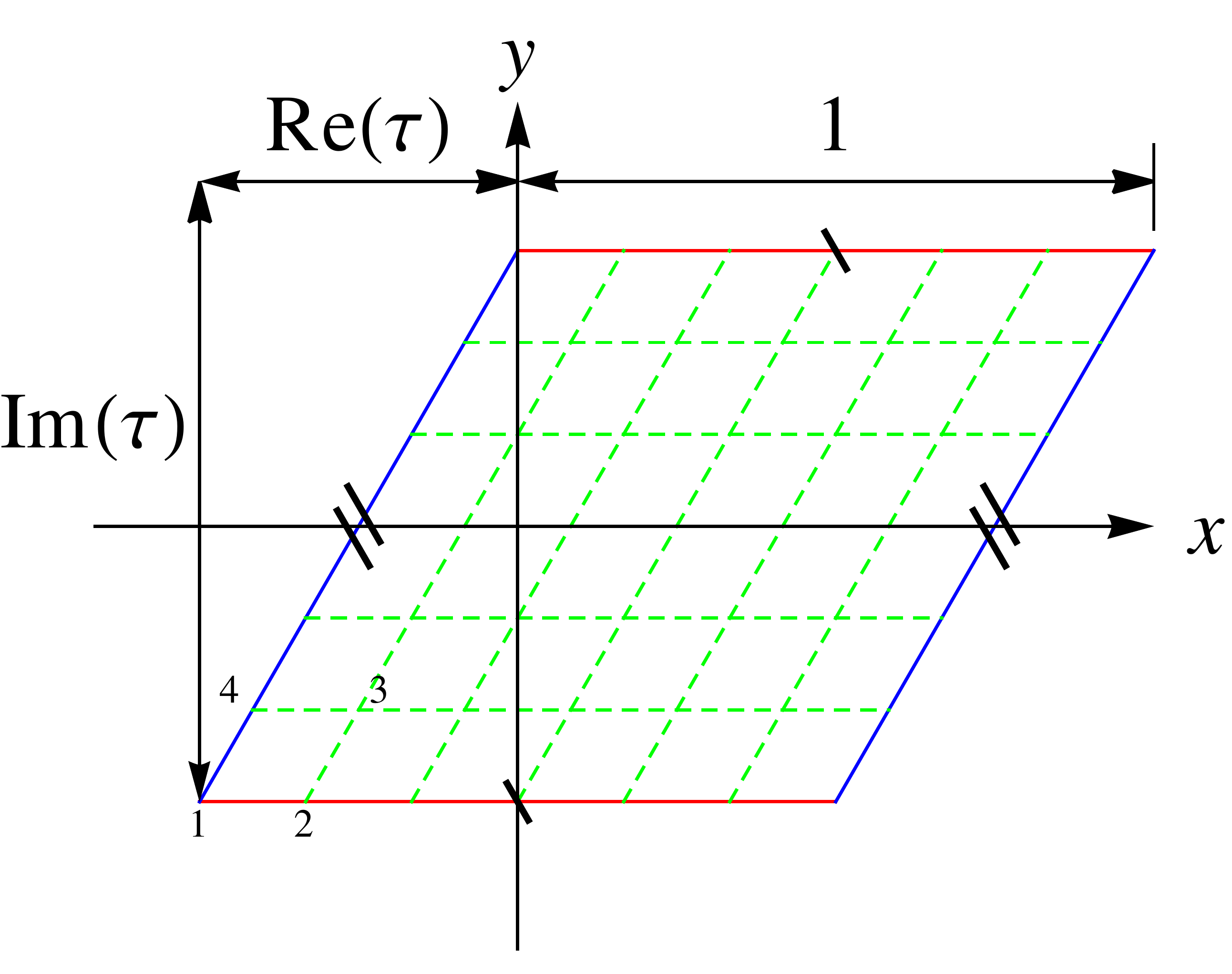}
  \caption{The lattice system can be put on a torus by imposing the equivalence conditions: $z\sim z+1$ and $z\sim z+\tau$ where $\tau=\tau_x+i\tau_y$ is a complex number. The principal region of a torus is bounded by the four points $z=\frac{1}{2}(\pm1\pm\tau)$. Here the top and bottom, left and right sides are identified, respectively.}
 \label{fig:torus}
\end{figure}

The GPWF can be put on a torus by implying the equivalence conditions: $z\sim z+1$ and $z\sim z+\tau$, as shown in Fig. \ref{fig:torus}. The principal region of a torus is bounded by the four points $z=\frac{1}{2}(\pm1\pm\tau)$. The torus is defined by two primitive vectors $\vec{\omega}_1=1$ and $\vec{\omega}_2=\tau_x+i\tau_y$. The shape of the torus is invariant under the $SL(2,\mathbb{Z})$ transformations $\begin{pmatrix}\vec{\omega}_1' \\\vec{\omega}_2'\end{pmatrix} = M\begin{pmatrix}\vec{\omega}_1 \\\vec{\omega}_2\end{pmatrix}$ with $M\in SL(2,\mathbb{Z})$ and the generators ($\hat{S}$ and $\hat{T}$) have the expressions
\begin{eqnarray}\label{eq:ST}
  \hat{S}=\begin{pmatrix} 0&-1\\1&0\end{pmatrix},\quad \hat{T}=\begin{pmatrix}1 &1\\0 &1\end{pmatrix}.
\end{eqnarray}

Two different constructions of GPWF for a CSL in the lattice analogy of $\nu=\frac{1}{2}$ bosonic Laughlin state can be found in Refs. \onlinecite{Kalmeyer1987,Wen1989a}. In Ref. \onlinecite{Kalmeyer1987}, the parton wave functions are discretized integer quantum Hall states and we call it \emph{ideal} GPWF for a CSL. On a torus, we can explicitly write down the ideal GPWF in terms the Laughlin-Jastrow wave functions\cite{Haldane1985}
\begin{eqnarray}
\label{eq:idealGPWF}  &&\mathcal{P}_G\Psi(\{z_i^\uparrow,z_k^\downarrow\})=e^{i\frac{K^\uparrow-K^\downarrow}{2}(Z^\uparrow-Z^\downarrow)}\nonumber\\
&\times&\vartheta_{\frac{1}{2},\frac{1}{2}}(Z^\uparrow-Z_0^\uparrow|\tau)\vartheta_{\frac{1}{2},\frac{1}{2}}(Z^\downarrow-Z_0^\downarrow|\tau)\nonumber\\
&\times&\mathcal{P}_G\prod_{i<j}^{N^\uparrow}\vartheta_{\frac{1}{2},\frac{1}{2}}(z_i^\uparrow-z_j^\uparrow|\tau)\prod_{k<l}^{N^\downarrow}\vartheta_{\frac{1}{2},\frac{1}{2}}(z_k^\downarrow-z_l^\downarrow|\tau),
\end{eqnarray}
where $\theta_{a,b}(z|\tau)$ is the theta function and $Z^\sigma=\sum_iz_i^\sigma$ is the center-of-mass coordinate. Different ground states are specified by the different zeros, $Z_0^\sigma$, in the center-of-mass wave functions. The zeros are determined by the general boundary conditions.\cite{Haldane1985,Niu1985} The modular $S$ and $T$ matrices for the ideal GPWF in Eq.(\ref{eq:idealGPWF}) can be analytically calculated by deformation the mass matrix\cite{Wen2012}
\begin{eqnarray}
\label{eq:STIdeal}
  S=\frac{1}{\sqrt{2}}\begin{pmatrix}1 &1\\1 &-1\end{pmatrix},  T=e^{-i\frac{2\pi c}{24}}\begin{pmatrix}1 &0\\0 &e^{i\frac{\pi}{2}}\end{pmatrix}.
\end{eqnarray}
with the central charge $c=1$, the same as those for the $\nu=\frac{1}{2}$ bosonic Laughlin state.

In Ref. \onlinecite{Wen1989a},  the \emph{general} GPWF for a CSL is written as
\begin{eqnarray}
\label{eq:generalGPWF}  \mathcal{P}_G\Psi(\{z_i^\uparrow,z_i^\downarrow\})=\mathcal{P}_G\det\varphi_i(z_j^\uparrow)\det\varphi_k(z_l^\downarrow),
\end{eqnarray}
where $\det\varphi_i(z_j^\uparrow)$ is the determinate wave function for the fermionic partons filling the valence bands of the tight binding model
\begin{eqnarray}
\label{eq:TBM}
  H_{\text{MF}}=-\sum_{ij,\sigma}t(z_i,z_j) f_{\sigma}^\dag(z_i) f_{\sigma}(z_j)+\text{H.C.},
\end{eqnarray}
on the $\pi$-flux square lattice with both nearest neighbor and next nearest neighbor hopping amplitude.\cite{Wen1989a} There are $\frac{\pi}{2}$ flux in every triangle in the plaqutte, e.g. $\triangle_{123}$ in $\square_{1234}$ in Fig. \ref{fig:torus}, $\Phi(\triangle_{123})=\arg(t_{z_1z_2}t_{z_2z_3}t_{z_3z_1})=\frac{\pi}{2}$. Different ground state wave functions can be obtained by different general boundary conditions. For the spin operator, the boundary condition is
\begin{eqnarray}
\label{eq:SO}
  S^+(z_i+1)=e^{i\Phi^s_1}S^+(z_i),~ S^+(z_i+\tau)=e^{i\Phi^s_2}S^+(z_i).\nonumber
\end{eqnarray}
Due to fractionalization in the GPWF\cite{Mei2014,Liu2014}, the parton has the boundary condition
\begin{eqnarray}
  f_{\sigma}^\dag(z_i+1)=e^{i\frac{\sigma}{2}\Phi^s_1}f_{\sigma}^\dag(z_i),\quad f_{\sigma}^\dag(z_i+\tau)=e^{i\frac{\sigma}{2}\Phi_2^s}f_{\sigma}^\dag(z_i),\nonumber
\end{eqnarray}
with $\sigma=\pm1$ for $f_{\uparrow/\downarrow}^\dag$. When we increase $\Phi_{1,2}^s$ from $0$ to $2\pi$, the spin operators is invariant, however, the parton wave functions do not go back to themselves and lead to another ground state for GPWF. Therefore, we have different ground states for a CSL labeled by the spin fluxes in the holes of a torus $|\Phi_1^s,\Phi_2^s\rangle$,
\begin{eqnarray}
\label{eq:statewithflux}
  \{|\Psi_a\rangle\}=\{|0,0\rangle, |0,2\pi\rangle, |2\pi,0\rangle,  |2\pi,2\pi\rangle\},
\end{eqnarray}
with $a=1,2,3,4$. Actually only two of them are linearly independent.

For the general GPWF in Eq. (\ref{eq:generalGPWF}), we use the UWFO in Eq. ({\ref{eq:uwfo}}) to exact the modular matrices $S$ and $T$. To carry out the UWFO, we need calculate the following overlaps
\begin{eqnarray}\label{eq:matrices}
  P_{ab}=\langle\Psi_a|\Psi_b\rangle,~\tilde{S}_{ab}=\langle\Psi_a|\Psi_b^{S}\rangle,~ \tilde{T}_{ab}=\langle\Psi_a|\Psi_b^{T}\rangle.
\end{eqnarray}
where $|\Psi_a\rangle$ is the sate in Eq. (\ref{eq:statewithflux}) and $
|\Psi_b^S\rangle=\hat{S}|\Psi_b\rangle$, $|\Psi_b^T\rangle=\hat{T}|\Psi_b\rangle$,
where $\hat{S}$ and $\hat{T}$ are the 90\degree rotation and Dehn twist transformations in Eq. (\ref{eq:ST}) on a torus. The $P$ matrix has rank 2 with the numerical tolerance less than $10^{-3}$ implying two-fold ground state degeneracy.

Given GPWFs, we implement the ``sign trick''\cite{Zhang2011} to calculate the overlap
\begin{eqnarray}
\langle\Psi_a|\Psi_b\rangle&=&\sum_{\{z_i\}}\psi_a^*(\{z_i\})\psi_b(\{z_i\})\nonumber\\
 &\equiv&\langle\Psi_a|\Psi_b\rangle_\text{Amp}\langle\Psi_a|\Psi_b\rangle_\text{Sign}
\end{eqnarray}
where $\psi_a(\{z_i\})$ is the amplitude wave function of the spin configuration $\{z_i\}$ in $|\Psi_a\rangle$ and the sign term
\begin{eqnarray}
   \langle\Psi_a|\Psi_b\rangle_\text{Sign}=\sum_{\{z_i\}}\rho_{ab}\frac{\psi_i^*(\{z_i\})\psi_j(\{z_i\}}{|\psi_i(\{z_i\})\psi_j(\{z_i\})|}
\end{eqnarray}
is calculated by Monte Carlo method according to the weight $\rho_{ij}=|\psi_i(\{z_i\})\psi_j(\{z_i\})|$. The amplitude term is the normalization factor for weight $\rho_{ab}$
\begin{eqnarray}
  \langle\Psi_a|\Psi_b\rangle_\text{Amp}&=&\sum_{\{z_i\}}|\psi_a(\{z_i\})\psi_b(\{z_i\})|.
\end{eqnarray}
Actually, we are only interested in the ratios of amplitudes. For example, for $P$ matrix in Eq. (\ref{eq:matrices}), we evaluate the matrix-element amplitude ratios
\begin{eqnarray}
 \label{eq:ampRatio} \frac{\langle\Psi_a|\Psi_b\rangle_\text{Amp}}{\langle\Psi_1|\Psi_1\rangle_\text{Amp}}=\frac{\sum_{\{z_i\}}\rho_{ab;11}\sqrt{|\frac{\psi_a(\{z_i\})\psi_b(\{z_i\})}{\psi_1(\{z_i\})\psi_1(\{z_i\})}|}}{\sum_{\{z_i\}}\rho_{ab;11}\sqrt{|\frac{\psi_1(\{z_i\})\psi_1(\{z_i\})}{\psi_a(\{z_i\})\psi_b(\{z_i\})}|}}
\end{eqnarray}
according to the Monte Carlo sampling weight $\rho_{ab;11}=\sqrt{|\psi_i(\{z_a\})\psi_b(\{z_i\})\psi_1(\{z_i\})\psi_1(\{z_i\})|}$.

\begin{figure}[t]
  \centering
  \includegraphics[width=0.8\columnwidth]{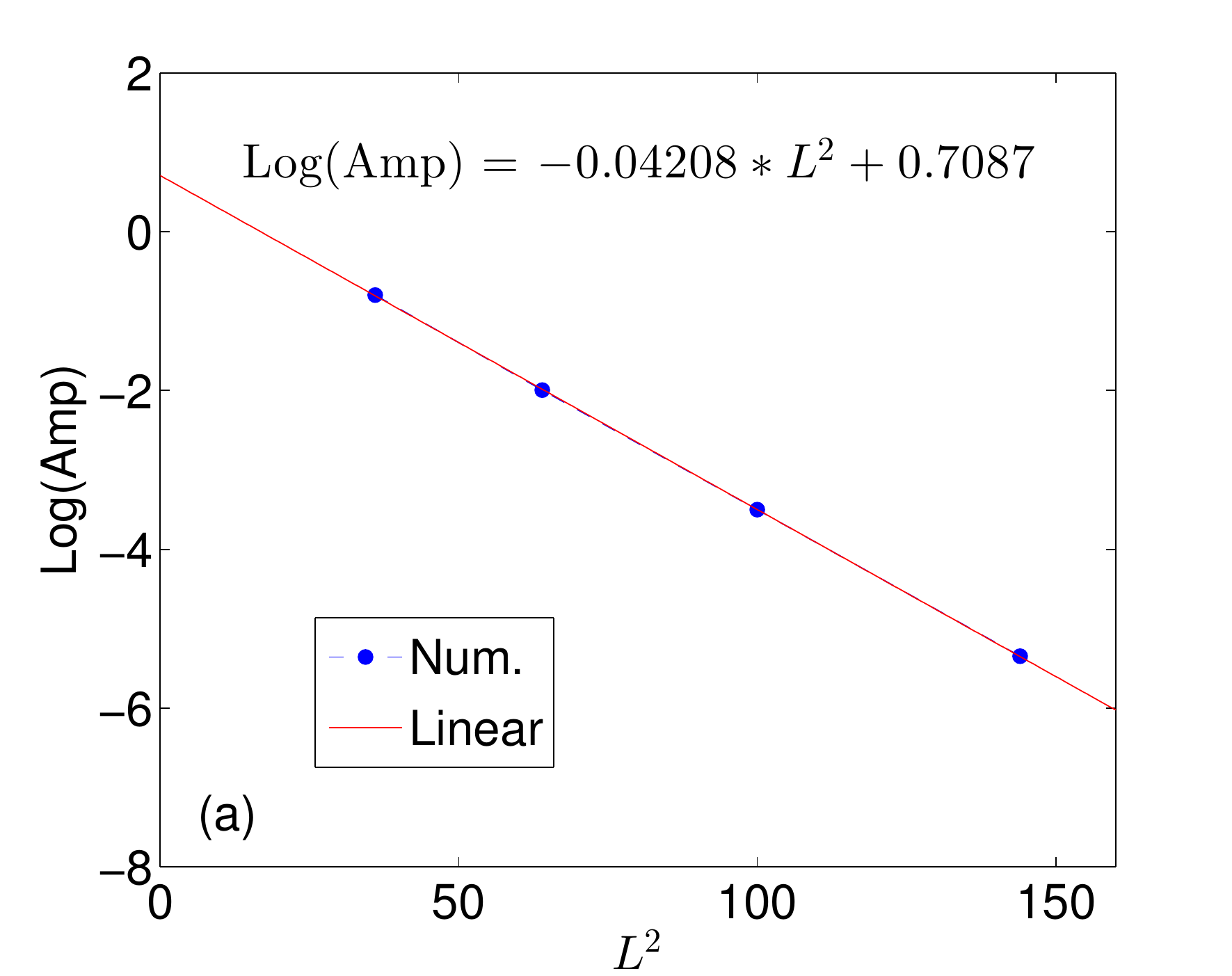}\\
  \includegraphics[width=0.8\columnwidth]{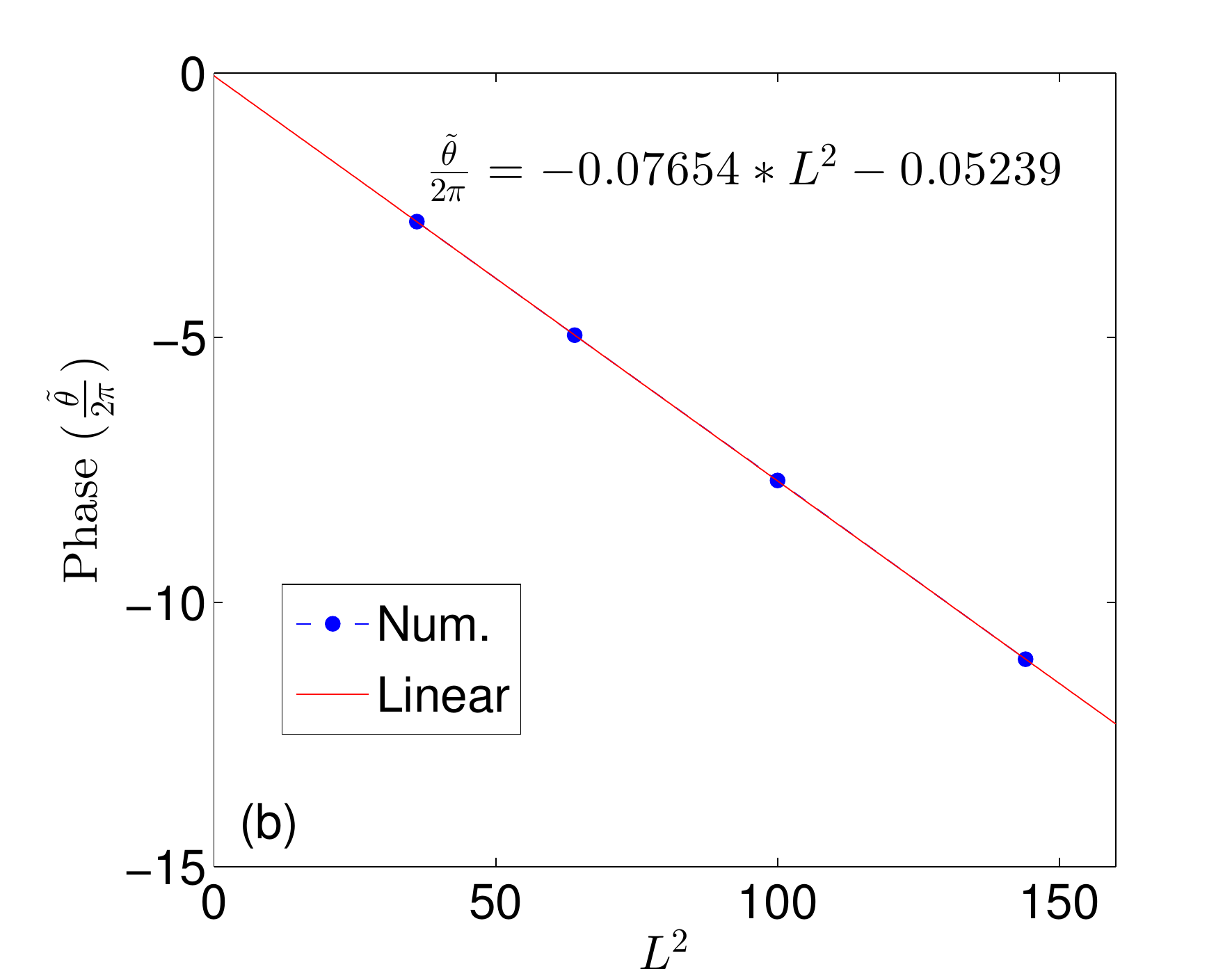}\\
  \includegraphics[width=0.8\columnwidth]{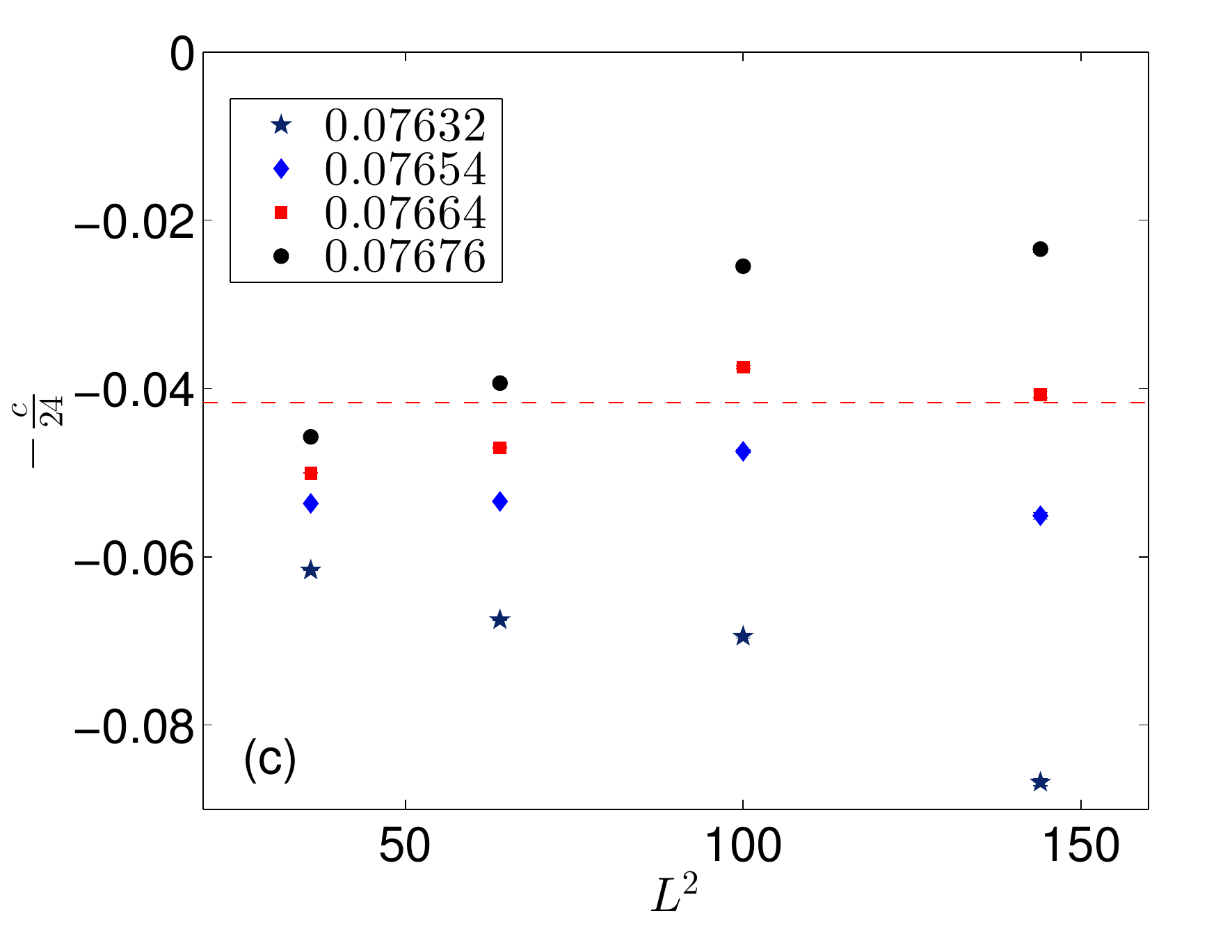}
  \caption{$L^2$-dependent of amplitude and phase of $T'$ in Eq. (\ref{eq:ST1}) are shown in (a) and (b), respectively. Here $\text{Log(Amp)}\equiv\log(|T'_{11}|)$ and $\frac{\tilde{\theta}}{2\pi}\equiv\frac{\arg{T'_{11}}}{2\pi}+k$ with $k=3,5,8,11$ for $L=6,8,10,12$. In (c), we plot $-\frac{c}{24}=\frac{\arg(T_{11})}{2\pi}+\frac{\text{Im}(\alpha_T)}{2\pi}L^2 \mod 1$ with different $\frac{\text{Im}(\alpha_T)}{2\pi}=0.07632,0.07654,0.07664,0.07676$. The red dashed line is for $c=1$. In (c), the numerical error bars are included and smaller than the symbols' sizes.  }
  \label{fig:ST}
\end{figure}
We set the mean field hopping parameters as $t_1/t_0=0.5$, where $t_0$ and $t_1$ are for nearest neighbor and next nearest neighbor links, respectively. The overlap calculations are carried out on the systems with $L\times L$ lattice sizes, $L=6,8,10,12$. From the overlaps in Eq. (\ref{eq:matrices}), we follow the steps below to extract the modular $S$ and $T$ matrices. We first digonalize the $P$ matrix
\begin{eqnarray}
  \label{eq:3}
  P=U^\dag P_\Lambda U, \quad U=(u_1,u_2,u_3,u_4).
\end{eqnarray}
Only two eigenvectors (e.g. $u_3$ and $u_4$) have non-zero eigenvalues around $2$. These two states ($u_3$ and $u_4$) are the linearly independent ground sates. In terms of the normalized $\tilde{U}=(u_3,u_4)$, the overlaps for $\tilde{S}$ and $\tilde{T}$ in Eq. ({\ref{eq:matrices}}) turn out to be $2\times2$ square matrices
\begin{eqnarray}
   S_{2\times2}^1=\tilde{U}^\dag\tilde{S}_{4\times4}\tilde{U},\quad   T_{2\times2}^1=\tilde{U}^\dag\tilde{T}_{4\times4}\tilde{U}.
\end{eqnarray}
Generally, $T^1$ is not diagonal since $u_3$ and $u_4$ are not the minimum entangled states or eigenstates of the Wilson loop operators\cite{Zhang2012}. We then diagonalize $T_1$ to obtain the minimum entangled states $v_1$ and $v_2$
\begin{eqnarray}\label{eq:ST1}
   T^1=V^\dag T' V, \quad S^1=V^\dag S' V, \quad V=(v_1,v_2),
\end{eqnarray}
where $T$ is diagonal and the phases of $V$ are fixed according to the conditions $S'_{12}=S'_{21}$ and $S'_{1i}>0$.

Since the CSL wave function has the 90\degree ~rotation symmetry, the exponent in $S'$ in Eq.(\ref{eq:ST1}) vanishes, $\alpha_S=0$, that is confirmed in the numerical calculations. The UWFO of the $T$ matrix has a complex exponent $\alpha_T$ in the prefactor. The real part of the exponent $\text{Re}(\alpha_T)$ is easily obtained from the amplitude of the $T'$ in Eq.(\ref{eq:ST1}) by fitting $\text{Log(Amp)}\equiv\log(|T'_{11}|)$ with respect to $L^2$, $\text{Re}(\alpha_T)=0.04208$, as shown in  Fig. \ref{fig:ST} (a). The phase $\tilde{\theta}$ is defined up $2\pi$, $\frac{\tilde{\theta}}{2\pi}\equiv\frac{\arg(T_{11}')}{2\pi}+k=-\frac{\text{Im}(\alpha_T)}{2\pi}L^2-\frac{c}{24}$ with $k\in \mathbb{Z}$. For $L=6,8,10,12$, the corresponding integers are $k=3,5,8,11$. From the fitting in Fig. \ref{fig:ST} (b), we obtain $\text{Im}(\alpha_T)=0.07654\times2\pi$. The central charge is sensitive to the exact value of $\text{Im}(\alpha_T)$ as shown in Fig. \ref{fig:ST} (c). The final result for the modular $S$ and $T$ matrices is
\begin{eqnarray}
  S=\begin{pmatrix}
   0.714&   0.707\\
   0.707&  -0.698
  \end{pmatrix},~ T=e^{-i\frac{2\pi c}{24}}\begin{pmatrix}
   1&  0\\
   0&  e^{i0.501\pi}
  \end{pmatrix},
\end{eqnarray}
with the central charge $c\simeq1.25\pm0.5$, very close to the exact result for the ideal GPWF in Eq. (\ref{eq:STIdeal}).

\begin{figure}[b]
  \centering
  \includegraphics[width=\columnwidth]{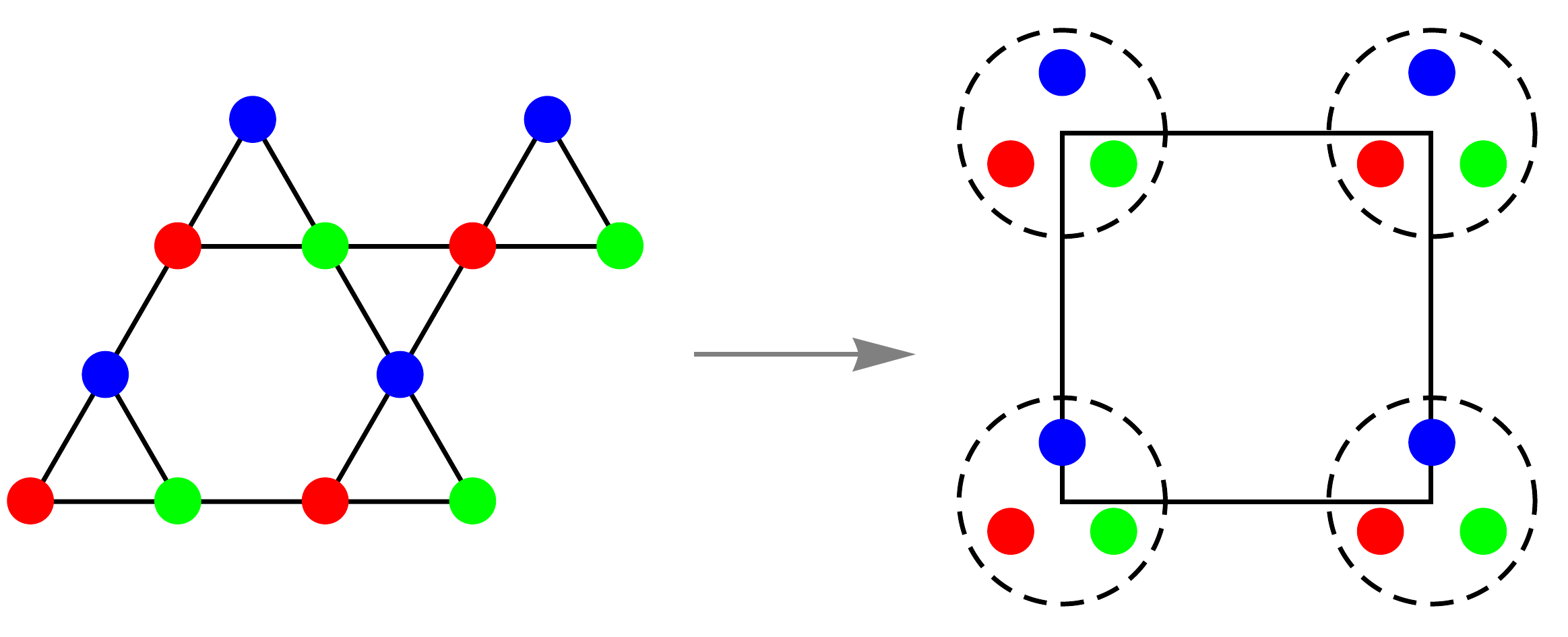}
  \caption{Kagome lattice is mapped onto a square lattice with three orbitals per site.}
  \label{fig:kagome}
\end{figure}

Above we apply the UWFO method on the square lattice. For a general Bravais
lattice, we can firstly map it onto an equivalent square lattice. We take the
kagome lattice as example. We map the unit cell of the kagome lattice onto the
one with square unit cell. Different sites within the unit cell are mapped
onto different orbitals on the square lattice, as shown in
Fig.\ref{fig:kagome}. Then we can make the modular transformations $\hat{S}$
and $\hat{T}$ on the square lattice torus. On the square lattice, we can also
use Kadanoff block renormalization procedure to reduce the system size
$L^2\rightarrow \tilde{L}^2$. Then the exponents in the prefactors of UWFO can
be significantly reduced. Many local unitary transformations on the lattice can
potentially reduce the exponents in the UWFO. If different ground state sectors
have the same topological spins, we can follow Ref. \onlinecite{Liu2013} to
identify the minimum entangled states to diagonalize the modular $T$ matrix.
The UWFO method is easily generalized to the 3+1D topological orders in the
GPWFs. The GPWF for quantum dimer models in 3D has already been constructed in
Ref. \onlinecite{Ivanov2014}. In 3D, the modular group of the 3-torus is
$SL(3,\mathbb{Z})$ generated by
\begin{eqnarray}
  \hat{S}=\begin{pmatrix}0 &1 & 0\\0&0&1\\1&0&0
  \end{pmatrix},\quad
  \hat{T}=\begin{pmatrix}1&0&0\\1&1&0\\0&0&1
  \end{pmatrix}.
\end{eqnarray}
We can use the UWFO to directly study the topological information in 3+1D\cite{Moradi2014}.

In conclusion, we use the universal wave function overlap method to exact the
modular $S$ and $T$ matrices for the topological order in the
Gutzwiller-projected parton wave function for the chiral spin liquid state on
the square lattice. The chiral spin liquid is the lattice analogy of
$\nu=\frac{1}{2}$ bosonic Laughlin state and the analogy is directly confirmed
by the modular $S$ and $T$ matrices from the universal wave function overlap.
The exponents in the prefactors of the wave function overlaps are found to be
small and the variational Monte Carlo calculations are carried out on
relatively large systems. The Monte Carlo calculations of the universal wave
function overlap can be easily generalize to other Bravais lattices and 3+1D
topological orders.

X-G. W is supported by NSF Grant No.  DMR-1005541 and NSFC 11274192.  He is
also supported by the BMO Financial Group and the John Templeton Foundation.
Research at Perimeter Institute is supported by the Government of Canada
through Industry Canada and by the Province of Ontario through the Ministry of
Research.

\bibliography{STGPWF}
\end{document}